\documentclass[12pt]{article}
\usepackage{graphicx}
\usepackage{subfigure}
\usepackage{color}
\newcommand{\beq}{\begin{equation}}
\newcommand{\eeq}{\end{equation}}
\newcommand{\bea}{\begin{eqnarray}}
\newcommand{\eea}{\end{eqnarray}}

\begin{document}

\begin{titlepage}
\begin{flushleft}
       \hfill                       FIT HE - 14-03 \\
       \hfill                       
\end{flushleft}

\begin{center}
   {\huge Accelerated quarks and energy loss    \\ 
   \vspace*{2mm}
 in confinement theory \vspace*{2mm}
}
\end{center}

\begin{center}

\vspace*{2mm}
{\large Kazuo Ghoroku${}^{a}$\footnote[1]{\tt gouroku@dontaku.fit.ac.jp},
${}^{}$
Kouki Kubo${}^{a}$\footnote[2]{\tt kkubo252@gmail.com},
%
}\\

\vspace*{2mm}
{${}^{a}$Fukuoka Institute of Technology, Wajiro, 
Higashi-ku} \\
{
Fukuoka 811-0295, Japan\\}

\vspace*{3mm}
\end{center}

\begin{abstract}
We study the energy loss rate (ELR) {of the accelerated quark 
in terms of the holographic models for the two different
motions, linear acceleration and uniform rotation. They are examined by two different
non-conformal models with confinement.}
We found in both models that
the value of ELR is bounded from below by the string tension
of the linear confinement potential between quark and anti-quark.
The lower bounds of ELR are independent of the types of the motion
of the quark. They are determined by the string tension
at the world sheet horizon of the model. 
{These results are obtained when the model has the diagonal background metric.}

\end{abstract}

\end{titlepage}

\pagebreak{}
\section{Introduction}

The radiation from accelerating quarks is basic and interesting problem
of the Yang-Mills gauge theory. It has a great variety of aspects according
to the motion of the object in various stage of the interaction including the
strong coupling regime. It is therefore very difficult to describe theoretically
the radiation from the accelerating object. The energy loss of the accelerated
quarks are determined by various different mechanisms of rich dynamics
of the Yang-Mills theory. The approach from the lattice simulation
is also inadequate to understand this time dependent phenomenon. 

On the other hand, the gauge/gravity duality helps us to tackle this problem.
This approach is very powerful in solving the dynamics in the strong coupling regime
of the gauge theory. 
A moving quark, which obeys the dynamics of 4D gauge theory with strong coupling
constant, can be described by a classical string solved in
5D gravity. This string ends on the boundary of 5D bulk and the end points correspond
to the positions of the quark and anti-quark. We can obtain informations about
the radiation from this quark when it is accelerated by some accidental force. 

The charged particles like quarks radiate when they are accelerated as is well-known.
This radiation is related to the world sheet horizon of the string corresponding
to this accelerated quark.
The ELR (Energy Loss Rate) 
of the quark due to this radiation is obtained in terms of the world-sheet momentum $\Pi^{\sigma}_{t}$ as follows  
\cite{Xiao:2008nr}-\cite{AliAkbari:2011ue},
\begin{equation}
\frac{dE}{dt}\equiv\Pi^{\sigma}_{t}\equiv\frac{\delta S}{\delta\left(\partial_{\sigma}X^{0}\right)}\ \ \ \mbox{(at world sheet horizon)},\label{eq:def-ene-loss-rate}
\end{equation}
where $X^{0}$ is the time-component of the string embedded function.
$S$ denotes the string action, and the horizon on the world sheet is defined by the zero-point of the induced metric for
the string 
such that $g_{\tau\tau}=0$. 
The formula (\ref{eq:def-ene-loss-rate}) has been given by Xiao \cite{Xiao:2008nr} by considering the energy
of the string behind the world sheet horizon as the lost energy due to the radiation.
Although this formula has been shown in the AdS case, 
we extend it to the other bulk background.
On the other hand, the ELR for uniform rotating
quark has been given in Ref. \cite{Athanasiou:2010pv} for ${\cal N}=4$ SYM theory 
and also for non-conformal and confining D4/$S^1$ model in Ref. \cite{AliAkbari:2011ue}. 
In these cases, the ELR is given as a coordinate independent form, so the above
formula is also useful. 

\vspace{.3cm}
In the Ref. \cite{AliAkbari:2011ue}, it is pointed out that there is 
a lower bound of the ELR for the confinement case, and then the radiation is related to 
the jet of glueballs which have the mass gap given by the confinement mass scale.
In this model, we could find the same lower bound of ELR for the case of the uniform accelerating quark.  
Here we extend this analysis to the other non-conformal and confining Yang-Mills theory
\cite{Kehagias:1999iy,Liu:1999fc} for the uniform accelerating 
and the uniform rotating quarks with a constant angular velocity.

Then we could find the ELR lower bound being common to the both types of acceleration. 
This is not accidental. The lower bound is given by the tension, 
which is denoted as $\tau_{\rm QCD}$, of the linear potential
between the quark and anti-quark in both models. {This fact could be understood as follows.
In order to keep the accelerated motion of the quark considered here, we need an external force $F$.
It should overcome at least the force coming from the linear potential which supports the
quark confinement. So we consider as $F\geq \tau_{\rm QCD}$.} Furthermore, this force would provide the energy
to the quark and the stretching string.
The radiation part of the provided energy
would be evaluated at the world sheet horizon as the string energy 
elongated by the tension at this point, $\tau(z)$, within a unit time as
\beq\label{lowest}
   {dE\over dt}=\tau(z)v|_{\rm at ~horizon}=\tau(z)|_{\rm at ~horizon}\geq \tau_{\rm QCD}\, ,
\eeq
where $v(=1)$ denotes the velocity of the string at the world sheet horizon. 
The details are given below for each accelerated motion in the Sec. 3.

\vspace{.3cm}
The outline of this paper is as follows.
In the next section, the solutions of
supergravity, dAdS and D4/$S^1$,
are given as the dual of two SYM theories with confinement. In the Sec.3,
the ELRs for linear acceleration and uniform rotation are given for dAdS model. 
In Sec. 5, the same analysis is given for D4/$S^1$. In both cases, we could find
the lower bound of the ELR as shown by (\ref{lowest}).
The summary and discussions are given in the final section.

\section{Model}

In this work, we {consider the case of diagonal }background metric $G_{MN}$. 
Then we can write the action and the canonical momentum density associated with the string as
\begin{eqnarray*}
S=-\frac{1}{2\pi\alpha^{\prime}}\int d\tau d\sigma\sqrt{-g_{\tau\tau}g_{\sigma\sigma}+g_{\tau\sigma}^{2}},
\end{eqnarray*}
\begin{eqnarray*}
\Pi_{t}^{\sigma} = -\frac{1}{2\pi\alpha^{\prime}}\frac{-g_{\tau\tau}\partial_{\sigma}X^{0}G_{tt}+g_{\tau\sigma}\partial_{\tau}X^{0}G_{tt}}{\sqrt{-g_{\tau\tau}g_{\sigma\sigma}+g_{\tau\sigma}^{2}}},
\end{eqnarray*}
where $g_{ab}\equiv\partial_aX^M G_{MN} \partial_bX^N \  (a,b=\tau,\sigma)$ is the induced metric and $X^M(\tau,\sigma)$ is the string embedding function.
Choosing the gauge $\tau=t\equiv X^{0}$, we have the ELR which
defined by Eq. (\ref{eq:def-ene-loss-rate}), 
\begin{equation}
\frac{dE}{dt}=\left.\frac{\left|G_{tt}\right|}{2\pi\alpha^{^{\prime}}}\right|_{z=\overline{z}}\ge \tau_{QCD},\label{eq:general-energy-loss-rate}
\end{equation}
where $\overline{z}$ denotes the {world sheet horizon, $g_{\tau\tau}(\bar{z})=0.$}
\footnote{Since $\tau_{QCD}=\frac{\min(\left| G_{tt}\right|)}{2\pi\alpha^{\prime}}$, the confinement property of the gauge theory (non-zero value of $\tau_{QCD}$) directory indicates the existence of the lower bound of the ELR. }

We examine following two different holographic confining gauge theory model.

\paragraph{deformed AdS by dilaton (dAdS model)}

One of the models we use is very similar to the AdS model but it is
not conformal \cite{Kehagias:1999iy,Liu:1999fc}. The metric is written as
\begin{eqnarray}
ds^{2} & = & G_{MN}dX^{M}dX^{N}\nonumber \\
 & = & e^{\Phi/2}\frac{L^{2}}{z^{2}}\left(-dt^{2}+d\vec{x}^{2}+dz^{2}\right),\ \ \ e^{\Phi}=1+qz^{4}\label{eq:metric-dAdS}
\end{eqnarray}
The difference from AdS space-time is the part including constant
$q$ which breaks the conformal symmetry. Then the theory gets the
typical energy scale, $q^{1/4}$, as the free parameter.

This parameter $q$ plays an important role in this model: It makes
the model confinement. We can see that the quark-anti quark potential
proportional to the distance between the quarks times $\sqrt{q}$.
Thus the model with any finite $q$ is confinement.

\paragraph{D4/$S^1$ model}

Another model is Witten's D4/$S^1$ model which is very well known \cite{Witten:1998zw}. The model
constructed by $N$ D4 branes on a circle, and it gives the metric as
follows,
\begin{equation}
ds^{2}=\left(\frac{u}{l}\right)^{3/2}\left(-dt^{2}+d\vec{x}^{2}+fdx_{4}^{2}\right)+\left(\frac{l}{u}\right)^{3/2}\left(\frac{du^{2}}{f}+u^{2}d\Omega_{4}^{2}\right),\ \ f=1-\left(\frac{u_{k}}{u}\right)^{3},\label{eq:metric-D4}
\end{equation}
where $l$ is the typical length scale of this model.

\section{dAdS model}

\subsection{Linear acceleration(uniform acceleration)}

\begin{figure}
\begin{centering}
\includegraphics[scale=0.8]{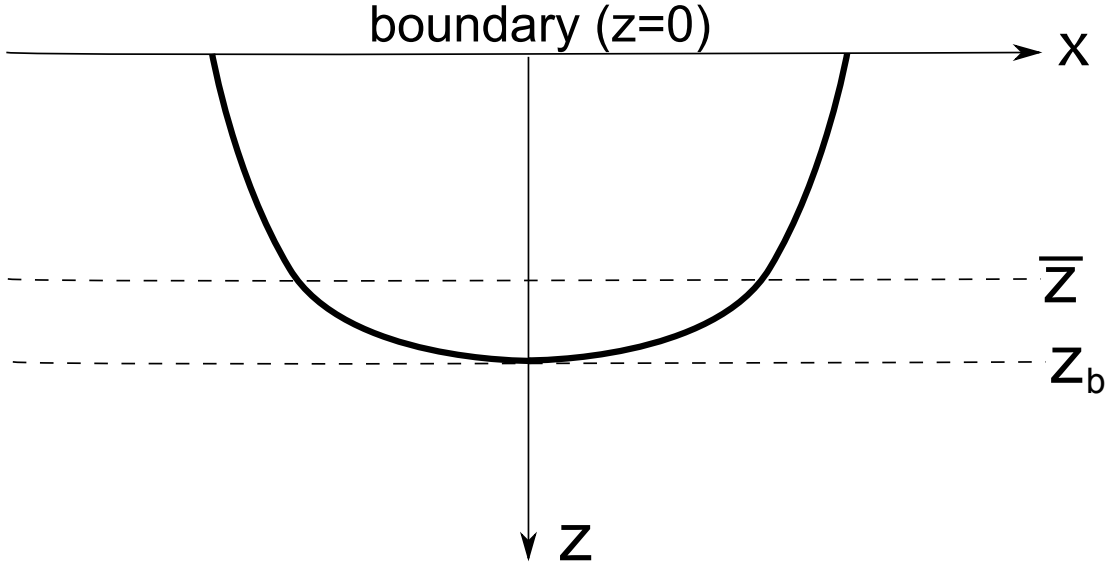}
\par\end{centering}

\caption{\label{fig:linearly-accelerating}The conceptual sketch of the linearly
accelerating string. There is an induced metric horizon at $z=\overline{z}$.
This string can be regarded as the pair of the quark and the antiquark.}

\end{figure}

First we consider the uniform accelerating quark in dAdS model. We
parameterize the string embedding function as
\[
X^{M}=\left(t,x\left(t,z\right),0,0,z\right).
\]
Here we choose $(\tau,\sigma)=(t,z)$.
Then the action and the equation of motion are written as  \cite{Ghoroku:2010sp}
\[
S=-\frac{1}{2\pi\alpha^{\prime}}\int dtdz\ e^{\Phi/2}\frac{L^{2}}{z^{2}}\sqrt{1-\dot{x}^{2}+x^{\prime2}},
\]
\begin{equation}
\partial_{t}\left(e^{\Phi/2}\frac{L^{2}}{z^{2}}\frac{-\dot{x}}{\sqrt{1-\dot{x}^{2}+x^{\prime2}}}\right)+\partial_{z}\left(e^{\Phi/2}\frac{L^{2}}{z^{2}}\frac{x^{\prime}}{\sqrt{1-\dot{x}^{2}+x^{\prime2}}}\right)=0.\label{eq:eom-uniform-acceleration}
\end{equation}
If we assume that the function $x$ can be written as%
\footnote{We can identify $g\left(0\right)$ with the reciprocal of the square
of the acceleration of the quark, namely,
\[
g\left(0\right)=\frac{1}{a^{2}}
\]
}
\beq
x=\sqrt{t^{2}+g\left(z\right)}, \label{profile}
\eeq
the equation (\ref{eq:eom-uniform-acceleration}) is reduced to 
\[
g^{\prime\prime}-2\left(\frac{1}{4}g^{\prime2}+\left(\frac{1}{z}\frac{1}{1+qz^{4}}g^{\prime}+1\right)\left(g+\frac{1}{4}g^{\prime2}\right)\right)\frac{1}{g}=0,
\]
and the induced metric on the world sheet becomes
\begin{eqnarray*}
g_{ab} & = & e^{\Phi/2}\frac{L^{2}}{z^{2}}\left(\begin{array}{cc}
\frac{-g\left(z\right)}{x^{2}} & \frac{1}{2}\frac{tg^{\prime}}{x^{2}}\\
\frac{1}{2}\frac{tg^{\prime}}{x^{2}} & \frac{t^{2}+g+\frac{1}{4}g^{\prime2}}{x^{2}}
\end{array}\right).
\end{eqnarray*}
The conceptual sketch of this string expressed by (\ref{profile}) is shown in Fig. \ref{fig:linearly-accelerating}.

\begin{figure}
\begin{centering}
\includegraphics[scale=0.8]{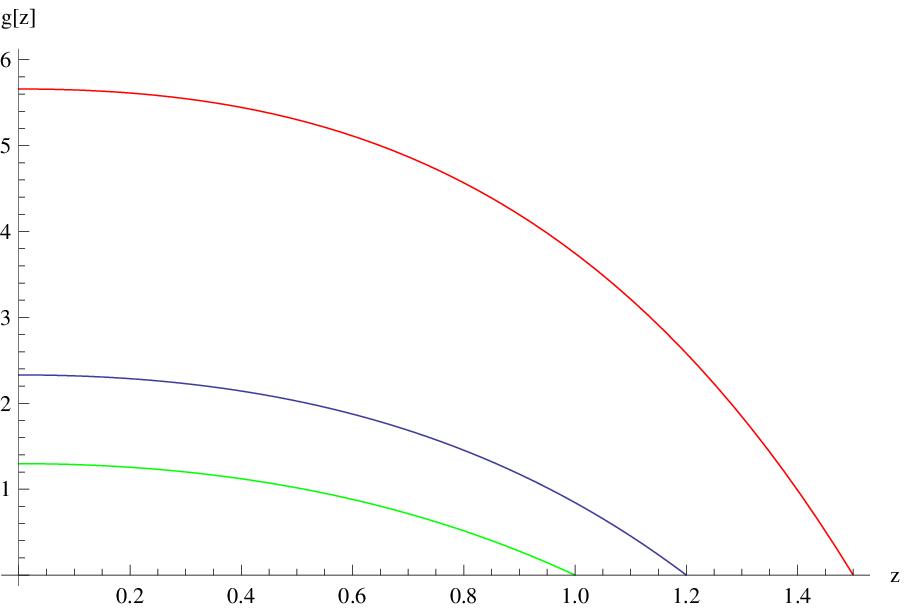}\includegraphics[scale=0.8]{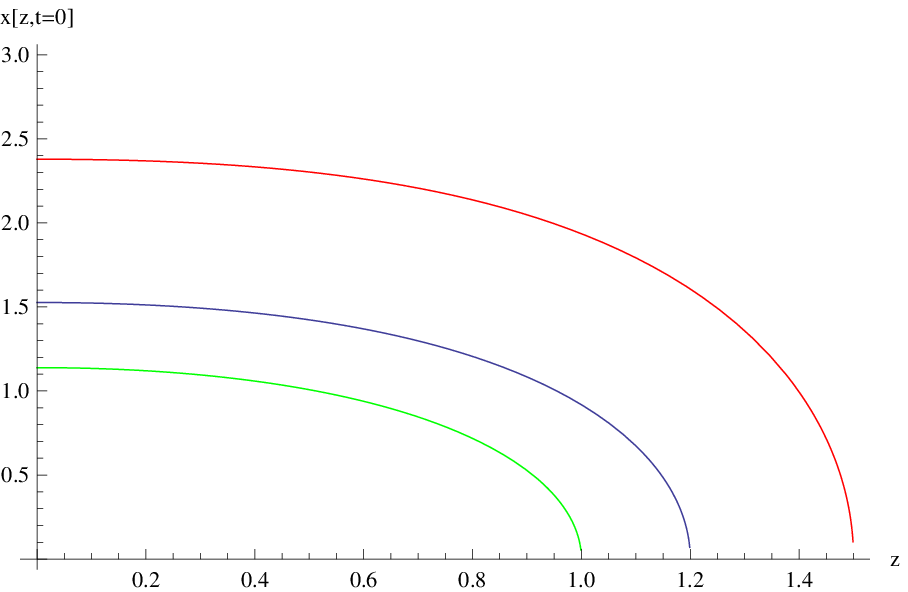}
\par\end{centering}

\caption{\label{fig:Numerical-plot-of-f}Numerical plot of $g\left(z\right)$ and
$x\left(z,t=0\right)$. ($\overline{z}=1,1.2,1.5$) It is clear that
$g\left(0\right)=1/a^2$ increases with $\overline{z}$. }

\end{figure}

The ELR is
\begin{eqnarray}
\frac{dE}{dt} = \frac{1}{2\pi\alpha^{\prime}}\frac{L^{2}}{\overline{z}^{2}}\sqrt{1+q\overline{z}^{4}}\ge\frac{L^{2}\sqrt{q}}{2\pi\alpha^{\prime}}.\label{eq:constraint-uniform}
\end{eqnarray}
There is the lower bound of this quantity, $\frac{L^{2}\sqrt{q}}{2\pi\alpha^{\prime}}$,
in the limit $\overline{z}\to\infty$. Since the horizon $\overline{z}$
goes far away in this limit, the limit corresponds to the small acceleration
limit. It can be seen in Fig. \ref{fig:Numerical-plot-of-f}. 

We notice here that the above lowest value of ELR is equivalent to the tension parameter
$\tau_{\rm QCD}(=\frac{L^{2}\sqrt{q}}{2\pi\alpha^{\prime}})$ \cite{GY}, 
which is defined by the quark and anti-quark potential $V$ as 
$V=\tau_{\rm QCD}r$ for large $r$, where $r$ denotes the distance between the quark and anti-quark.
This is expected as mentioned above through equation (\ref{lowest}).

\vspace{.5cm}
The above equation (\ref{eq:constraint-uniform}) is obtained by the consideration given in the introduction.
The ELR is evaluated as the energy of the part pulled by the tension at $z=\bar{z}$ for a unit time in the direction
of $x$. It is given as
\beq
  {dE\over dt}=\frac{1}{2\pi\alpha^{\prime}}\int_0^1 dx
\frac{L^{2}}{\overline{z}^{2}}\sqrt{1+q\overline{z}^{4}}
=\frac{1}{2\pi\alpha^{\prime}}
\frac{L^{2}}{\overline{z}^{2}}\sqrt{1+q\overline{z}^{4}}\, .
\eeq 
From this, we find the force given in (\ref{lowest}) as the tension at the horizon. The lowest value
of this tension is equivalent to $\tau_{\rm QCD}$.

\subsection{Uniform rotation}
Another interesting motion of the quark is the uniform rotation. 
According to \cite{Athanasiou:2010pv,AliAkbari:2011ue}, here
we use the cylindrical coordinates, $\tilde{X}^{M}=\left(t,\phi,\rho,x^{3},z\right)$,
namely, we transform $3$-space coordinates $\left(x^{1},x^{2},x^{3}\right)$
to the cylindrical coordinates $\left(\phi,\rho,x^{3}\right)$ as
\[
x^{1}=\rho\cos\phi,\ \ \ x^{2}=\rho\sin\phi,\ \ \ x^{3}=x^{3}.
\]

In the cylindrical coordinates system, the metric (\ref{eq:metric-dAdS})
is rewritten as
\begin{eqnarray*}
ds^{2} & = & e^{\Phi/2}\frac{L^{2}}{z^{2}}\left(-dt^{2}+d\rho^{2}+\rho^{2}d\phi^{2}+\left(dx^{3}\right)^{2}+dz^{2}\right),\ \ \ e^{\Phi}=1+qz^{4}.
\end{eqnarray*}
The ansatz for a rotating string is given by
\begin{eqnarray*}
\tilde{X}^{M} & = & \left(t,\phi,\rho,x^{3},z\right)\\
 & = & \left(t,\omega t+\theta\left(z\right),\rho\left(z\right),0,z\right).
\end{eqnarray*}
Then the induced metric $g_{ab}$ and the action $S_{rot}$ become

\begin{equation}
g_{ab}=\left(\begin{array}{cc}
g_{tt} & g_{tz}\\
g_{zt} & g_{zz}
\end{array}\right)=e^{\Phi/2}\frac{L^{2}}{z^{2}}\left(\begin{array}{cc}
-1+\omega^{2}\rho^{2} & \omega\theta^{\prime}\rho^{2}\\
\omega\theta^{\prime}\rho^{2} & 1+\theta^{\prime2}\rho^{2}+\rho^{\prime2}
\end{array}\right),\label{eq:metric-rotating}
\end{equation}
\begin{equation}
S_{rot}=-\frac{1}{2\pi\alpha^{\prime}}\int dtdz\ e^{\Phi/2}\frac{L^{2}}{z^{2}}\sqrt{\left(1-\omega^{2}\rho^{2}\right)\left(1+\rho^{\prime2}\right)+\theta^{\prime2}\rho^{2}},\label{eq:action-rotating}
\end{equation}
where the prime ($\prime$) denotes $\frac{\partial}{\partial z}$.

The equations of motion become
\begin{equation}
\partial_{z}\left\{ e^{\frac{\Phi}{2}}\frac{L^{2}}{z^{2}}\frac{\left(1-\omega^{2}\rho^{2}\right)\partial_{z}\rho}{\sqrt{\left(1-\omega^{2}\rho^{2}\right)\left(1+\left(\partial_{z}\rho\right)^{2}\right)+\rho^{2}\left(\partial_{z}\theta\right)^{2}}}\right\} =e^{\frac{\Phi}{2}}\frac{L^{2}}{z^{2}}\frac{-\omega^{2}\rho\left(1+\left(\partial_{z}\rho\right)^{2}\right)+\rho\left(\partial_{z}\theta\right)^{2}}{\sqrt{\left(1-\omega^{2}\rho^{2}\right)\left(1+\left(\partial_{z}\rho\right)^{2}\right)+\rho^{2}\left(\partial_{z}\theta\right)^{2}}},\label{eq:eom-rho}
\end{equation}
for $\rho$,
\begin{equation}
e^{\frac{\Phi}{2}}\frac{L^{2}}{z^{2}}\frac{\rho^{2}\partial_{z}\theta}{\sqrt{1-\omega^{2}\rho^{2}+\left(1-\omega^{2}\rho^{2}\right)\left(\partial_{z}\rho\right)^{2}+\rho^{2}\left(\partial_{z}\theta\right)^{2}}}=\Pi,\label{eq:eom-phi}
\end{equation}
for $\phi$, where $\Pi$ is the integration constant.

\paragraph{Solution}

We can easily solve Eq. (\ref{eq:eom-phi}) respect to $\partial_{z}\theta$.
The solution is
\begin{equation}
\partial_{z}\theta=\frac{1}{\rho}\sqrt{\frac{\Pi^{2}\left(1-\omega^{2}\rho^{2}\right)\left(1+\left(\partial_{z}\rho\right)^{2}\right)}{\left(e^{\frac{\Phi}{2}}\frac{1}{z^{2}}\right)^{2}\rho^{2}-\Pi^{2}}},\label{eq:sol-thetaprime}
\end{equation}
The fact that Eq. (\ref{eq:sol-thetaprime}) must be real requires
the following conditions,
\begin{eqnarray}
\rho\left(\overline{z}\right) & = & \frac{1}{\omega},\label{eq:bdy-cond-1}\\
\frac{1}{\Pi\omega} & = & \frac{\overline{z}^{2}}{e^{\frac{\Phi}{2}}}.\label{eq:cindition-zbar2}
\end{eqnarray}
Since $\left(tt\right)$-component of the induced metric (\ref{eq:metric-rotating})
is
\begin{eqnarray*}
g_{tt} & = & e^{\frac{\Phi}{2}}\frac{L^{2}}{z^{2}}\left(-1+\omega^{2}\rho^{2}\right),
\end{eqnarray*}
$z=\overline{z}$ agree with the position of the induced metric horizon.
Solving the condition (\ref{eq:cindition-zbar2}), we have
\begin{equation}
\overline{z}^{4}=\frac{1}{\Pi^{2}\omega^{2}-q}>0.\label{eq:rel-z-pi-q}
\end{equation}
This value must be positive because $\overline{z}$ is real number.
The condition gives a constraint among the parameters, $\Pi,\omega$
and $q$,
\begin{equation}
\Pi^{2}\omega^{2}-q>0.\label{eq:constraint-for-q-Piomega}
\end{equation}

Substituting the equation (\ref{eq:sol-thetaprime}) into the equation
of motion (\ref{eq:eom-rho}),we have the differential equation,
\begin{equation}
\partial_{z}^{2}\rho+\left(\frac{z\left(1+qz^{4}\right)\rho+2\rho^{2}\partial_{z}\rho}{z\left(z^{4}\Pi^{2}-\left(1+qz^{4}\right)\rho^{2}\right)}+\frac{1}{\rho}\frac{1}{1-\omega^{2}\rho^{2}}\right)\left(1+\left(\partial_{z}\rho\right)^{2}\right)=0.\label{eq:eom-rho-final-form}
\end{equation}

We can solve this equation numerically if we fix the boundary
condition. However, we cannot choose boundary conditions freely because
of the conditions (\ref{eq:cindition-zbar2}). To understand how the
boundary condition is determined, we expand the function $\rho\left(z\right)$
at $z=\overline{z}$,
\begin{equation}
\rho=\rho\left(\overline{z}\right)+\rho^{\prime}\left(\overline{z}\right)\left(z-\overline{z}\right)+\cdots.\label{eq:expand-rho}
\end{equation}
Substitute (\ref{eq:expand-rho}) into (\ref{eq:eom-rho-final-form})
and solving the differential equation perturbatively, we have
\begin{eqnarray}
\rho^{\prime}\left(\overline{z}\right) & = & \frac{\sqrt{\overline{z}^{2}\omega^{2}\left(1+q\overline{z}^{4}\right)^{2}+4}-\overline{z}\omega\left(1+q\overline{z}^{4}\right)}{2}.\label{eq:bdy-cond-2}
\end{eqnarray}
So we must choose the boundary conditions as (\ref{eq:bdy-cond-1})
and (\ref{eq:bdy-cond-2}). 

\begin{figure}
\begin{centering}
\includegraphics[scale=0.7]{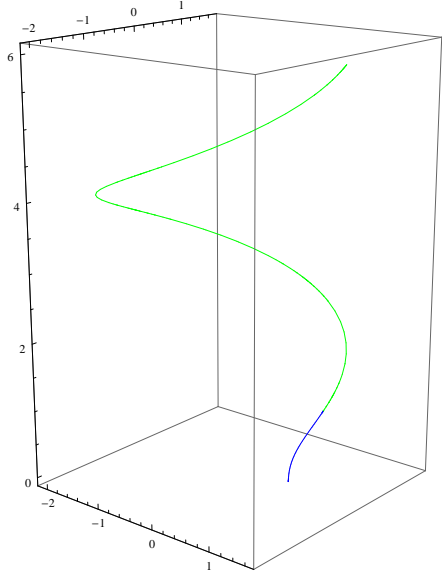}
\par\end{centering}

\caption{\label{fig:rotting-string}A numerical plot of  typical string configuration
with $\Pi=1$, $\omega=1$ and $q=0.1$.
 The lower [Blue] (upper [green]) line represents
the outside (inside) of the induced metric horizon. The boundary is
at $z=0$ (the bottom of this graph.)  The out side (the lower [blue]) part of
the string is connected to the boundary causally but inside (the upper [green])
part is not. Therefore the energy flow to the inside of the induced
metric horizon (the upper [green] part of the string) is identified with the ELR due to the radiation.}
\end{figure}

\begin{figure}
\begin{centering}
\includegraphics[scale=0.7]{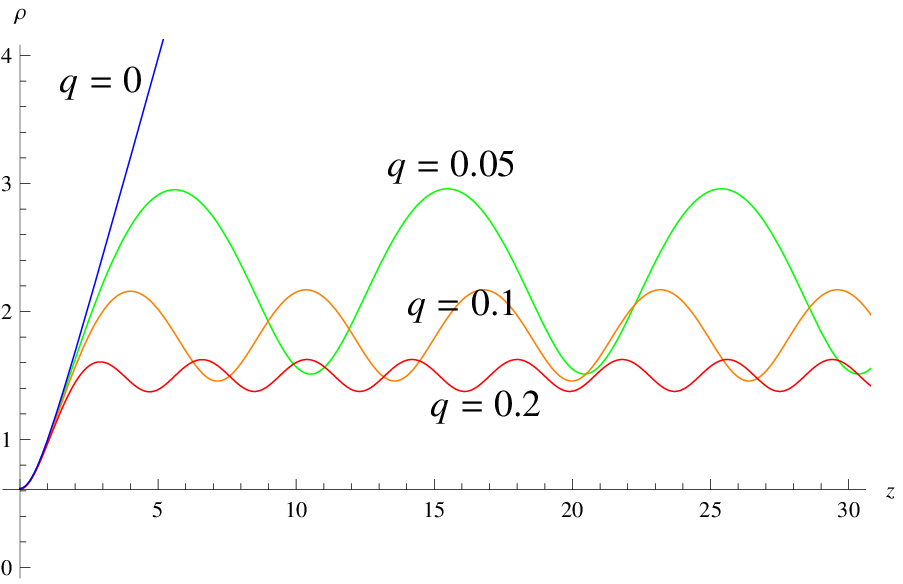}\includegraphics[scale=0.7]{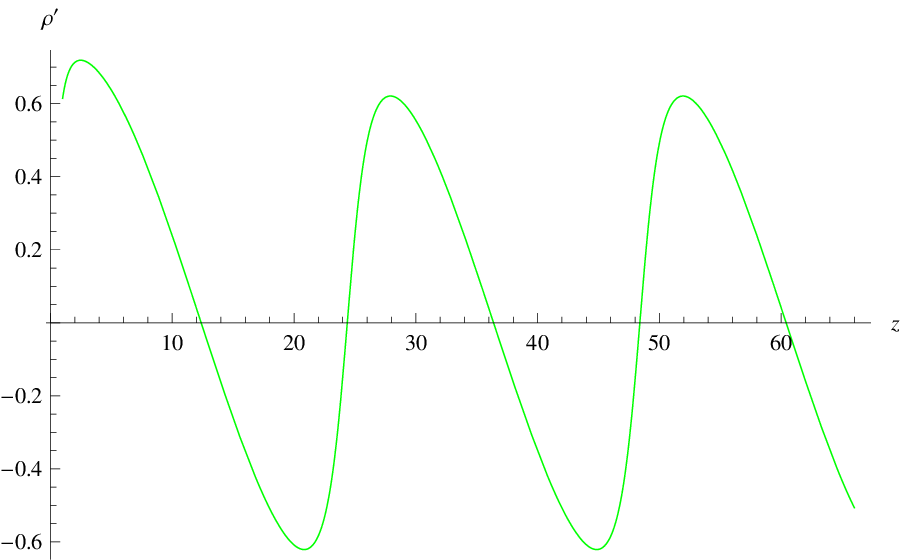}
\par\end{centering}

\caption{\label{fig:rho-z}Numerical plots of  typical string configurations
with $\Pi=1$, $\omega=1$. In the left, numerical plots of  $\rho(z)$ with  $q=0$ (blue), $q=0.05$ (green), $q=0.1$ (orange) and $q=0.2$ (red) are displayed.
The period and amplitude decrease with increasing $q$.
In the limit of $q \to 0$, the period and  amplitude diverge and the solution approaches to that of the AdS.
In the right, a numerical plot of $\rho^\prime (z)$ with $q=0.01$ is shown. We recognize that this oscillation cannot be described by simple trigonometric functions.
}
\end{figure}

A typical example of the solution of the string configuration is
shown in Fig. \ref{fig:rotting-string}.
The shape of the string in large $z$ region with finite $q$ is very different
 from that with $q=0$ \cite{Athanasiou:2010pv}. 
In AdS case ($q=0$), $\rho(z)$ is proportional to $z$ in large $z$ limit, $\rho(z)\sim z$, and it is divergent linearly.  In dAdS case ($q\ne 0$),  $\rho(z)$ seems to be oscillating around a certain positive value with a finite amplitude keeping $\rho>1/\omega$. (See Fig. \ref{fig:rho-z}.)
We can find following feature in Fig. \ref{fig:rho-z}.
\begin{itemize}
\item {For finite value of $q$, $\rho\left(z\right)$ oscillates with $z$.}
The period and the amplitude of the oscillation of $\rho\left(z\right)$
decrease with increasing $q$. 
\item The oscillation is not a simple trigonometric function.
\item We cannot compute the solution in large value of $q$ because of our
computer performance. However the solutions cross the induced metric
horizon $\left(\rho=1/\omega\right)$ only once in our calculations.
\item In $q\to0$ limit, we can confirm that the solution with finite $q$
agree with that with $q=0$. 
\end{itemize}


\vspace{.3cm}
The ELR for the uniform rotating quark is given by
\begin{eqnarray}
\frac{dE}{dt} & = & \frac{1}{2\pi\alpha^{\prime}}\frac{L^{2}}{\overline{z}^{2}}\sqrt{1+q\overline{z}^{4}}=\frac{\omega\Pi L^{2}}{2\pi\alpha^{\prime}}.\label{eq:energy-loss-rate-rotation-dAdS}
\end{eqnarray}
The lower bound arises from the constraint (\ref{eq:constraint-for-q-Piomega}),
\begin{eqnarray}
\frac{dE}{dt} & > & \frac{\sqrt{q}L^{2}}{2\pi\alpha^{\prime}}.\label{eq:constraint-rotating}
\end{eqnarray}
It is very interesting that the lower bound (\ref{eq:constraint-rotating})
agree with the value of the lower bound of the ELR for
uniform accelerating quark (\ref{eq:constraint-uniform}) 
in spite of the difference
of the motions.

\vspace{.5cm}
{In this case also, we can understand this lower bound as $\tau_{\rm QCD}$ 
from the minimum of the necessary force to pull the string for a unit time at the world sheet horizon. }
In the present case, the string can not be stretched to the centripetal
force to keep the rotation of the quark as mentioned in the introduction. The string at the horizon
should be pulled along the circle. As in the previous case, by using Eq. (\ref{eq:action-rotating}), we find
\beq
  \frac{dE}{dt}= \frac{1}{2\pi\alpha^{\prime}}\int_{0}^{\omega} d\theta~\rho
           \ e^{\Phi/2}\frac{L^{2}}{z^{2}}
           = \frac{1}{2\pi\alpha^{\prime}}\ e^{\Phi/2}\frac{L^{2}}{z^{2}}>\frac{\sqrt{q}L^{2}}{2\pi\alpha^{\prime}}\, ,
\eeq
where $\rho\omega=1$ and $z=\bar{z}$. Then we find again the same result with (\ref{eq:energy-loss-rate-rotation-dAdS})

\section{D4/$S^1$ model}

\subsection{Linear acceleration(uniform acceleration)}

We also investigate D4/$S^1$ model. The string embedding function is parameterized
as,
\[
X^{M}=\left(t,x_{1},x_{2},x_{3},x_{4},u,\Omega_{4}\right)=\left(t,x\left(t,u\right),0,0,0,u,\vec{0}\right).
\]
Then induced metric $g_{ab}$, string action $S$ and $\Pi_{u}^{t}$
become
\begin{equation}
g_{ab}=\left(\begin{array}{cc}
-\left(\frac{u}{l}\right)^{3/2}\left(1-\dot{x}^{2}\right) & \left(\frac{u}{l}\right)^{3/2}\dot{x}x^{\prime}\\
\left(\frac{u}{l}\right)^{3/2}\dot{x}x^{\prime} & \left(\frac{u}{l}\right)^{3/2}x^{\prime2}+\left(\frac{l}{u}\right)^{3/2}f^{-1}
\end{array}\right),\label{eq:induced-metric-linear-D4}
\end{equation}
\[
S=-\frac{1}{2\pi\alpha^{\prime}}\sqrt{\left(\frac{u}{l}\right)^{3}x^{\prime2}+f^{-1}-\dot{x}^{2}f^{-1}},
\]
and
\[
\Pi_{t}^{u}=\frac{1}{2\pi\alpha^{\prime}}\frac{\left(\frac{u}{l}\right)^{3}\dot{x}x^{\prime}}{\sqrt{\left(\frac{u}{l}\right)^{3}x^{\prime2}+f^{-1}-\dot{x}^{2}f^{-1}}}.
\]
From Eq. (\ref{eq:induced-metric-linear-D4}), we realize that the
induced metric horizon is found at the points $\left(\overline{t},\overline{u}\right)$
such that $\dot{x}\left(\overline{t},\overline{u}\right)^{2}=1$.
Thus ELR becomes
\begin{equation}
\frac{dE}{dt}\equiv\Pi_{t}^{u}\left(\overline{t},\overline{u}\right)=\frac{1}{2\pi\alpha^{\prime}}\left(\frac{\overline{u}}{l}\right)^{3/2}>\frac{1}{2\pi\alpha^{\prime}}\left(\frac{u_{k}}{l}\right)^{3/2}.\label{eq:lower-bound-linear-D4}
\end{equation}
Because $u>u_{k}$,$\frac{dE}{dt}$ has the lower bound. It is considerable
that the value of the lower bound (\ref{eq:lower-bound-linear-D4})
is coincide with the lower bound of the ELR of the rotating
string (\ref{eq:lower-bound-rotation-D4}).
{The situation is the same with the case of dAdS.
We can understand this lower bound as the minimum of the necessary force to overcome the string tension,
which is just $\frac{1}{2\pi\alpha^{\prime}}\left(\frac{u_{k}}{l}\right)^{3/2}=\tau_{\rm QCD}$ in this model. This is easily obtained
through the calculation of the Wilson loop.}

\subsection{Uniform rotation}

Because the detailed investigation of the rotating string in D4/$S^1$ model
is already given by Ref. \cite{AliAkbari:2011ue}. We just borrow
the result from the work,
\begin{equation}
\frac{dE}{dt}=\frac{1}{2\pi\alpha^{\prime}}\left(\frac{\overline{u}}{l}\right)^{3/2}>\frac{\Lambda_{QCD}^{3}}{2\pi\alpha^{\prime}}=\frac{1}{2\pi\alpha^{\prime}}\left(\frac{u_{k}}{l}\right)^{3/2}.\label{eq:lower-bound-rotation-D4}
\end{equation}
As expected, the ELR of the rotating quark has also the
lower bound which agrees with (\ref{eq:lower-bound-linear-D4}).

\section{Summary and discussion}

We find the common lower bound of the ELR for two different accelerated
motions of the quark, (\ref{eq:constraint-uniform}) and (\ref{eq:constraint-rotating})
or (\ref{eq:lower-bound-linear-D4}) and (\ref{eq:lower-bound-rotation-D4}).
The value is determined by the parameter $q$ or $u_{k}/l$. 
It is given precisely by $\tau_{\rm QCD}$ for each confinement model as expected

In dAdS model, the scale parameter $q$ controls the dynamics. 
This parameter appears in various physical quantity. 
If we take the limit $q\to 0$, the metric of this model (\ref{eq:metric-dAdS})
reduces to the one of the AdS space and the lower bound of the energy
loss late vanishes since $\tau_{\rm QCD}\to 0$.

In D4/$S^1$ model, we can find the same reason why the energy loss has the
lower bound. {The lower bound of ELR is given by the tension $\tau_{\rm QCD}$ of this model.
{Then, we may consider the present radiation as the jet of glueballs  
as pointed out in Ref. \cite{AliAkbari:2011ue}. This assertion can be supported also by our result.}


In both models, we found the common lower bounds of the ELR for two different accelerated
motions of the quark. {This point could be understood easily. In the models used here,
the bulk metric $G_{MN}$ is diagonal. Then, when a world sheet horizon exists in the 
embedded string configuration, the Eq. (\ref{eq:general-energy-loss-rate}) can be 
derived without any other assumption for $X^M(\tau,\sigma)$. This implies that 
the Eq. (\ref{eq:general-energy-loss-rate}) is available to various kinds of motion.
}
\footnote{
We notice that the similar equation
with (\ref{eq:general-energy-loss-rate}) is also found in Ref. \cite{Fadafan:2012qu}}
{Due to this fact, the results given by
Eqs. (\ref{eq:constraint-uniform}), (\ref{eq:lower-bound-linear-D4}), (\ref{eq:constraint-rotating})
, and (\ref{eq:lower-bound-rotation-D4}) are all understood from Eq. (\ref{eq:general-energy-loss-rate}).}

\section*{Acknowledgements}
The work of K.K. is supported by MEXT/JSPS,
Grant-in-Aid for JSPS Fellows No. 25$\cdot$4378.


\begin{thebibliography}{99}
\bibitem{Xiao:2008nr}
  B.~W.~Xiao,
  Phys.\ Lett.\ B {\bf 665} (2008) 173
  [arXiv:0804.1343 [hep-th]].

\bibitem{Ghoroku:2010sp}
  K.~Ghoroku, M.~Ishihara, K.~Kubo and T.~Taminato,
  Phys.\ Rev.\ D {\bf 83} (2011) 024020
  [arXiv:1010.4396 [hep-th]].

\bibitem{Athanasiou:2010pv}
  C.~Athanasiou, P.~M.~Chesler, H.~Liu, D.~Nickel and K.~Rajagopal,
  Phys.\ Rev.\ D {\bf 81} (2010) 126001
   [Phys.\ Rev.\ D {\bf 84} (2011) 069901]
  [arXiv:1001.3880 [hep-th]].

\bibitem{AliAkbari:2011ue}
  M.~Ali-Akbari and U.~Gursoy,
  JHEP {\bf 1201} (2012) 105
  [arXiv:1110.5881 [hep-th]].

\bibitem{Kehagias:1999iy}
  A.~Kehagias and K.~Sfetsos,
  Phys.\ Lett.\ B {\bf 456} (1999) 22
  [hep-th/9903109].

\bibitem{Liu:1999fc}
  H.~Liu and A.~A.~Tseytlin,
  Nucl.\ Phys.\ B {\bf 553} (1999) 231
  [hep-th/9903091].
\bibitem{GY} 
  K. Ghoroku and M. Yahiro, 
  Phys.\ Lett.\ B {\bf 604}, 235 (2004). 

\bibitem{Witten:1998zw}
  E.~Witten,
  Adv.\ Theor.\ Math.\ Phys.\  {\bf 2} (1998) 505
  [hep-th/9803131].

\bibitem{Fadafan:2012qu}
  K.~B.~Fadafan and H.~Soltanpanahi,
  JHEP {\bf 1210} (2012) 085
  [arXiv:1206.2271 [hep-th]].
\end{thebibliography}
\end{document}